\documentclass[aps,showpacs,twocolumn,superscriptaddress]{revtex4}
\usepackage{graphicx}
\usepackage{amsmath,amssymb}

\newcommand\ve[1]{\mathbf{#1}}
\newcommand\tu{\hat{\mathbf t}}

\begin{document}
\title{Hydrodynamic flow patterns and synchronization of beating cilia}
\author{Andrej Vilfan}
\email{Andrej.Vilfan@ijs.si}
\affiliation{J. Stefan Institute, Jamova 39, 1000 Ljubljana, Slovenia}
\author{Frank J\"ulicher}
\email{julicher@mpipks-dresden.mpg.de}
\affiliation{Max Planck Institute for the Physics of
   Complex Systems, N\"othnitzer Str. 38, 01187 Dresden, Germany}

\begin{abstract}
We calculate the hydrodynamic flow field generated far from a cilium which is
attached to a surface and beats periodically. In the case of two beating cilia,
hydrodynamic interactions can lead to synchronization of the cilia, which are
nonlinear oscillators.  We present a state diagram where synchronized states
occur as a function of distance of cilia and the relative orientation of their
beat.  Synchronized states occur with different relative phases.  In addition,
asynchronous solutions exist. Our work could be relevant for the synchronized
motion of cilia generating hydrodynamic flows on the surface of cells.
\end{abstract}

 \maketitle 

Many eucaryotic cells possess cilia, which are motile, whip-like structures on
the cell surface \cite{Bray2001,Brennen.Winet1977}. While certain cells such as
sperm swim using a single cilium (in this case called flagellum) other cells
use several or a large number of cilia for propulsion in a fluid.  Epithelial
cells such as those in the respiratory tract use densely ciliated surfaces to
transport fluids.  Hydrodynamic flows generated by cilia also play a key role
during the morphogenesis of higher organisms. In mammals, the left-right
symmetry of the embryo is broken with the help of nodal cilia which generate a
hydrodynamic flow to one side which transports signaling molecules and breaks
the symmetry
\cite{McGrath.Brueckner2003,Cartwright.Tuval2004,Okada.Hirokawa2005}.  All
these cilia are based on the same structure, called the axoneme, which is built
of a very regular arrangement of protein filaments called microtubules in a
cylindrical geometry. Motility is achieved by the action of a large number of
dynein motor proteins which generate forces in the cilium while consuming a
chemical fuel. As a consequence, the cilium produces periodic deformations in
three dimensional space.

\begin{figure}
  \begin{center}
    \includegraphics[width=8.5cm]{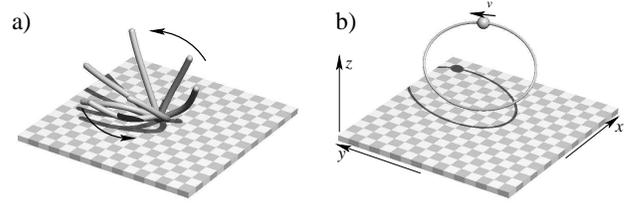}
  \end{center}
  \caption{a) Beating pattern of a cilium. 
  The effective stroke and the recovery stroke are different (arrows)
  and a net flow is generated. b) Simplified representation of  
  the cilium by a small sphere, moving on a tilted elliptical
    trajectory.} 
  \label{fig_model}
\end{figure}

Cilia in different organisms differ mainly by their length and their pattern of
beating. They typically measure few micrometers up to a few tens of micrometers
in length and around $150-300\,\rm nm$ in diameter.  While the beat of sperm
tails is typically a planar, sometimes helical wave, more complex, three
dimensional, asymmetric beating patterns occur typically in cells which propel
fluids along surfaces.  In this case, the working cycle of a cilium consists of
a fast, upright, effective stroke and a recovery stroke which brings the cilium
more slowly back to the original position on a path closer to the surface, see
Fig.~\ref{fig_model}a.  Nodal cilia break the left-right symmetry of developing
embryos by generating rotatory motion in a plane tilted with respect to the
surface to which they are attached \cite{Okada.Hirokawa2005}.

In this letter, we discuss the hydrodynamic flow field generated by a single
cilium which beats in an asymmetric pattern while attached to a surface.  While
the near field of the hydrodynamic flow depends on the detailed beating pattern
and the whip-like geometry of the cilium, the far field has general features
which only depend on a few parameters characterizing the symmetry of the beat.
Using the flow field generated by a beating cilium, we study the hydrodynamic
interaction between two cilia. We discuss conditions that lead to
synchronization of the two cilia by hydrodynamic interactions.

Our minimal model of the ciliary beat (Fig.~\ref{fig_model}b) captures the
essential features and symmetries - the difference between effective stroke
and recovery stroke of whip-like beating as well as the tilted rotatory motion 
of nodal cilia.
We replace the cilium by a small sphere of radius $a$
(essentially describing the center of mass position of the cilium), which moves
on a fixed trajectory in the vicinity of a planar surface (defined as the plane
$z=0$).  The trajectory of the bead is elliptic, the phase of the oscillation
(the position along the trajectory) is described by an angle $\phi$. The
asymmetry of the ciliary beat is reflected in the fact that the two principal
axes (denoted $A$ and $B$) are different, and that the ellipse is tilted with
respect to the surface as described by the parameter $C$.  The position of the
sphere is given by
\begin{equation}
  \label{eq:trajectoryi}
  \ve x_i=\ve x_i^0+ \ve x(\phi_i)={\left( \begin{array}{c} x_i^0\\y_i^0\\0 \end{array}
  \right)} +\left(\begin{array}{c}
A \cos \phi_i\\
B \sin \phi_i\\
D+C \cos \phi_i
\end{array}
\right)
\end{equation}
here, $x_i^0$ and $y_i^0$ describes the position on the surface at which the
cilium is attached.  For more detailed descriptions of the ciliary beat see,
e.g., refs.~\cite{Brennen.Winet1977,Gueron.Levit-Gurevich1998}.

The viscous, over damped fluid around the cilium can be described by the Stokes
equation for the hydrodynamic flow field ${\bf v}$
\begin{equation}
\eta \nabla^2 {\bf v}=\nabla P
\end{equation}  
where the pressure $P$ is a Lagrange multiplier to impose the constraint of
incompressibility, $\nabla\cdot{\bf v}=0$. The far-field generated by a moving
sphere in the vicinity of a surface can be calculated by studying the solution
of the Stokes equation to a delta-distributed force ${\bf F}_i$ at position
${\bf x}_i=(x_i,y_i,z_i)$ which is of the form $\ve v(\ve x)=\tensor G(\ve
x_i,\ve x) \ve F_i$.  The tensor $\tensor G$ consists of a stokeslet $\tensor
G^S$, which describes the velocity field around a small isolated particle, and
an image, required to satisfy the no-slip boundary condition on the surface
\cite{Blake.1971}.  The image is located at the particle position, mirrored
over the boundary plane, $\ve{\bar x}_i = \ve{x}_i - 2 z_i \hat e _z$, and
consists of an anti-stokeslet, a source-doublet $G^{D}$ and a stokes-doublet
$G^{SD}$, so that $\tensor G(\ve x_i,\ve x) = \tensor G^S (\ve{x}-\ve{x}_i) -
\tensor G^S (\ve{x}-\ve{\bar x}_i) + 2 z_i^2 \tensor G^D (\ve{x}-\ve{\bar x}_i)
-2z_i \tensor G^{SD} (\ve{x}-\ve{\bar x}_i)$ with
\begin{align}
  G_{\alpha \beta}^S(\ve{x})&= \frac 1 {8 \pi \eta}\left( \frac{\delta_{\alpha
  \beta}}{\left| \ve{x} \right|} +  \frac{\ve{x}_{\alpha}\ve{x}_{
  \beta}}{\left| \ve{x} \right|^3} \right) \label{eq:gs}\\
  G_{\alpha \beta}^D (\ve{x})&= \frac{1}{8 \pi \eta} (1-2\delta _{\beta z})
  \frac{\partial}{\partial
    x_\beta} \left( \frac{x_\alpha}{\left| \ve{x} \right|^3} \right)\\
  G_{\alpha \beta}^{SD} (\ve{x})&=(1-2\delta _{\beta z})
  \frac{\partial}{\partial x_\beta} G_{\alpha z}^S(\ve{x})\;. \label{eq:gsd}
\end{align}
 The leading behavior of  $G$ which describes the far-field 
 (for $r\gg z_i,z$) can be written as
\begin{equation}
  \label{eq:gapprox}
\tensor  G(\ve{x}_i,\ve{x})\approx \frac{3}{2 \pi \eta} \frac{z_i z}{r^3}
\left( \begin{matrix} \cos^2 \beta& \sin \beta \cos \beta & 0\\
\sin \beta \cos \beta & \sin ^2 \beta& 0\\
0 & 0& 0
\end{matrix}\right)
\end{equation}
where $\tan\beta=(y_i-y)/(x_i-x)$ and $r^2=(x_i-x)^2+(y_i-y)^2$.  For constant
height $z$, $G$ decays as as $G\propto r^{-3}$.

We assume that the active mechanism in the cilium generates a tangential force
${\bf f}_i=f_i \tu_i$. In a simplified model, it is a linear function of the
velocity ${\bf v}_i=v_i\tu_i$ described by $f_i=f_0-\kappa v_i$. Here, $\tu_i$
is a normalized vector parallel to ${\bf t}_i=d{\bf x}_i/d\phi_i$.  Note that
the total force ${\bf F}_i$ acting on the sphere is in general not parallel to
the direction of motion since normal forces arise to establish the constraint
which keeps the sphere moving on the ellipsoidal track.  The force ${\bf F}_i$
thus obeys
\begin{equation}
\label{eq:forcev}
  \tu_i \cdot {\bf F}_i = f_0- \kappa v_i\;.
\end{equation}
This force is balanced by hydrodynamic friction $\ve{F}_i=\tensor \gamma_i
\ve{v}_i$, where the friction matrix is given by \cite{Dufresne.Grier2000}
\begin{equation}
  \label{eq:friction-result}
\tensor{\gamma}_i= \tensor{\gamma}(\ve{x}_i)=\gamma_0\left(\tensor{I}+\frac {9}{16} \frac a {z_i} \left(
  \begin{array}{ccc}1&&\\&1&\\&&2\end{array} \right) \right) \; .
\end{equation}
Here $\gamma_0=6\pi \eta a$ denotes the Stokes friction of a small sphere with
radius $a$ and the second term the corrections due to proximity of the plane.
The force-velocity relation (\ref{eq:forcev}), balanced by the friction force
leads to the equation of motion for the phase of oscillation
\begin{equation}
  \label{eq:eqofm}
    \frac{d\phi_i}{dt}=f_0\left (\tu_i \cdot \tensor\gamma(\ve{x}_i)
   \ve{t}_i+\kappa \tu_i \cdot \ve{t}_i 
   \right)^{-1}
\end{equation}
The resulting motion of the sphere is periodic in time.  In the limit of small
radius $a$, the friction coefficient becomes independent of the height over the
surface and the period is $T=\frac{2\pi}{\omega}\simeq
\frac{(\kappa+\gamma_0)\ell}{f_0}$, where $\ell$ denotes the trajectory length.
The phase $\phi_0$ of the sphere as a function of time can be written as
\begin{equation}
  \label{eq:fourier}
  \phi_0 (\omega t)=\omega t + K \sin (\omega t) + L \sin (2 \omega t) + \ldots
\end{equation}
The coefficient $L$ is related to the eccentricity of the ellipse (if the
particle moves at a constant speed, the parameter $\phi$ has a variable time
derivative) and thus depends on the geometry of the trajectory.  The
coefficient $K \propto aC/D^2$, describes variations of the velocity due to
variations of the friction (\ref{eq:friction-result}) resulting form varying
distance of the particle from the surface.
 
\begin{figure}
  \begin{center}
    \includegraphics[scale=0.7]{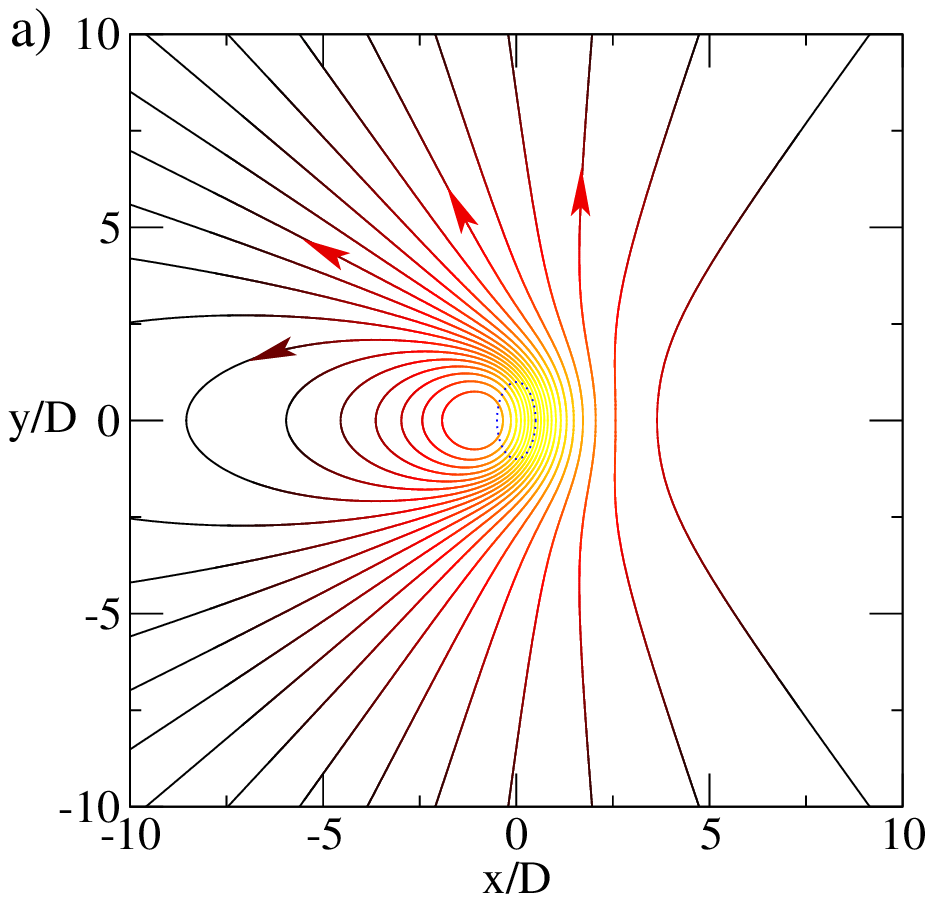}
    \includegraphics[scale=0.7]{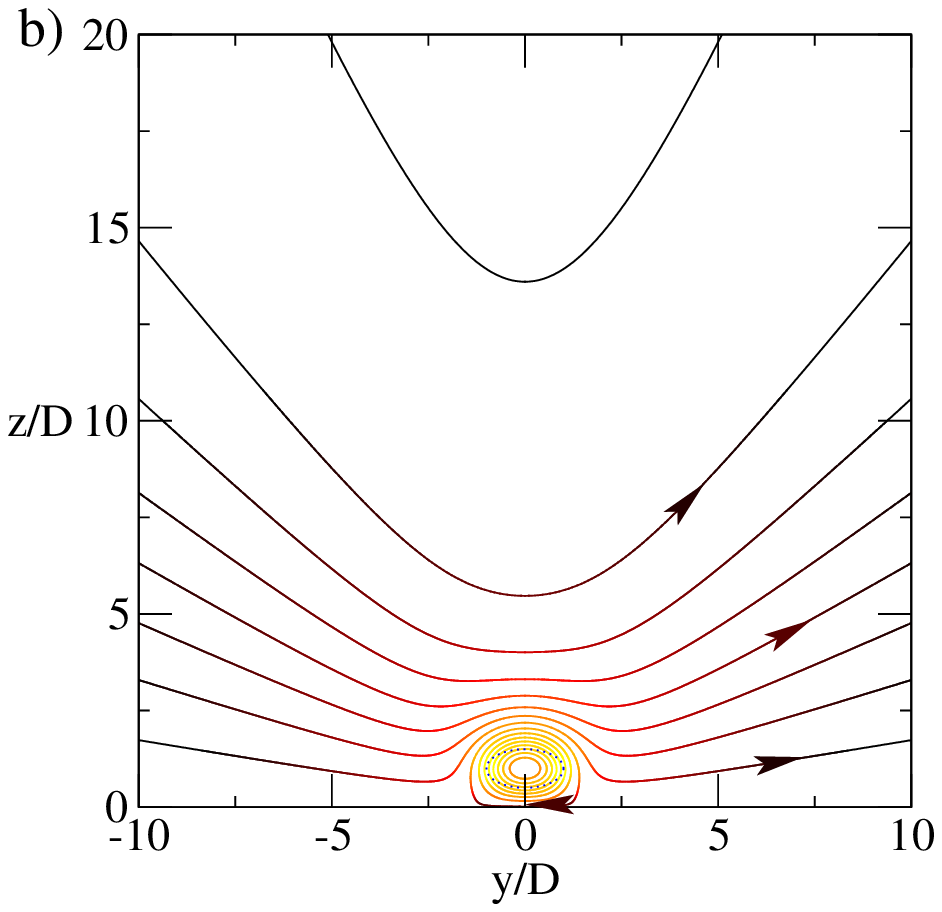}
  \end{center}
  \caption{a) Flux lines describing the time averaged fluid 
    flow generated by a sphere moving on a tilted elliptical path shown in the
    $x-y$ plane for $z/D=2$ (other parameters: $A/D=0.5$, $B/D=1$, $C/D=0.5$,
    $a/D=0.1$).  The curves were obtained numerically using the flow field
    based on Eqns. (\ref{eq:gs})-(\ref{eq:gsd}).  b) Same flow field displayed
    in the $y$-$z$ plane, for $x=0$. Black lines correspond to low fluid
    velocity, yellow lines to high velocity.  The dashed lines indicate the
    projections of the sphere trajectory.}
  \label{fig:fluxlines}
\end{figure}

We can now obtain the hydrodynamic far-field of a beating cilium ${\bf v}({\bf
  x},t)=\tensor G({\bf x}_i(t),{\bf x})\tensor \gamma_i (\ve{x}_i(t))
{\ve{v}}_i(t)$.  Averaging over one period, we define the net flow $\bar{\bf
  v}({\bf x}) =\frac{1}{T}\int_0^T \ve{v}(\ve{x},t)dt=\frac{1}{T}\oint \tensor
G({\bf x}_i,{\bf x})\tensor \gamma_i (\ve{x}_i)d \ve{x}_i $.  It only depends
on the period and the shape of the trajectory, but not on details in the phase
$\phi_0(t)$.  Because the sphere causes a stronger fluid motion when it is
further away from the surface (during the working stroke) than when it is
closer (recovery stroke), it causes a net fluid flow in $y$ direction.
Examples of flux lines of the time-averaged fluid flow generated in this way
are displayed in Fig.~\ref{fig:fluxlines}.  Far from a sphere moving around the
center at $x_i^0=0$, $y_i^0=0$, the average velocity field is given by
\begin{equation}
  \label{eq:farfieldv}
  \bar{\ve{v}}= \frac{9\pi a BC}{T}\frac{yz}{\left| \ve{x} \right|^5} \left(
  \begin{array}{c} x \\ y \\ z 
  \end{array} \right) 
\end{equation}
Note that it only depends on the sphere radius $a$, the period $T$ and the
projected area of the trajectory to the $y$-$z$ plane, $\pi BC$.  Corrections
to this flow field can be neglected in the far field since they decay at least
as ${\cal O}(|\ve{x}|^3)$ \cite{Lenz.Prost2003}.

\begin{figure}[htbp]
  \begin{center}
    \includegraphics{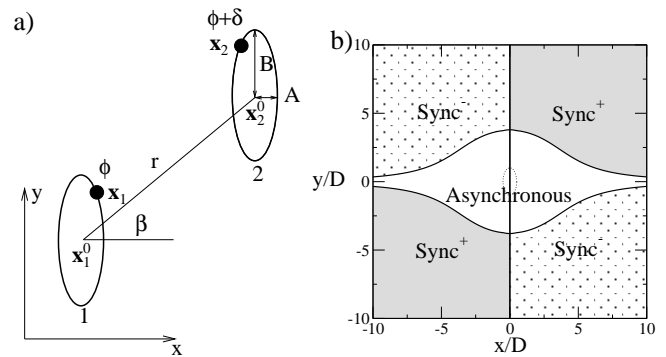}~\includegraphics[scale=0.65]{Figure3b.eps}\\
  \end{center}
  \caption{a) Top view of the arrangement of 
    two elliptical trajectories which represent two beating cilia. The distance
    $r$ and the angle $\beta$ are indicated. b) The state diagram as a function
    of the distance vector $x=x_2^0-x_1^0$, $y=y_2^0-y_1^0$, determined from
    the numerical solution of the full model equations. Three different regions
    are indicated. In the region of asynchronous beat, two frequencies occur.
    Two synchronous states can be distinguished with equal (Sync$^+$) and
    opposite (Sync$^-$) phase of both cilia in the limit of large separation
    $r$.}
  \label{fig_phasediag}
\end{figure}

\begin{figure}
  \begin{center}
    \includegraphics{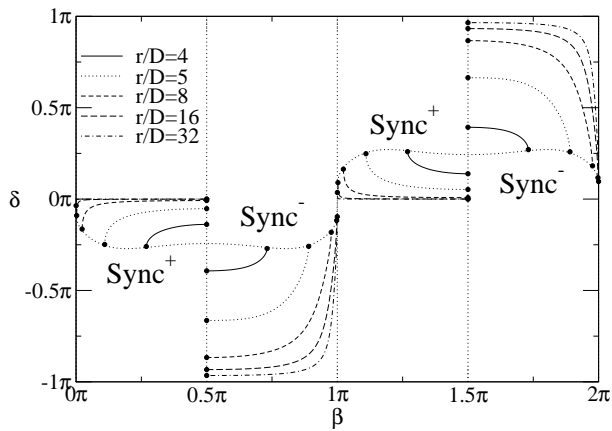}
  \end{center}
  \caption{The phase difference $\delta$ between two spheres on elliptic trajectories 
    in the stable stationary state as a function of their relative position in
    polar coordinates (angle $\beta$, distance $r$) obtained by numerical
    solutions to the full dynamic equations.  In the limit $r\to \infty$
    analytic far-field calculations show that only equal and opposite phases
    ($\delta=0$ and $\pm \pi$, respectively) occur.  For smaller $r$,
    nontrivial angles $\delta$ occur. Synchronous states become unstable where
    the lines end (black dots). Between these points motion is asynchronous.  }
  \label{fig_phasevsbeta}
\end{figure}

We now turn to the case of  two beating cilia which interact hydrodynamically.
The force ${\bf F}_2$ the second sphere exerts on the fluid at position ${\bf x}_2$
generates hydrodynamic flows at position ${\bf x_1}$ and thus influences
the force ${\bf F}_1$ exerted by the first sphere:
\begin{equation}
  \label{eq:friction0}
  {\ve{F}}_1=\tensor{\gamma}_1 \left( \ve{v}_1 -  \tensor{G}(\ve{x}_2,\ve{x}_1) \ve{F}_2 \right)
\end{equation}
This equation can be used to define the dynamics of motion of both spheres
along their ellipsoidal trajectories in the following manner.  We assume again
that each sphere is driven by a force obeying Eq.~(\ref{eq:forcev}).
Eq.~(\ref{eq:friction0}) together with the corresponding equation for ${\bf
  F}_2$ then uniquely determines the tangential velocities of both spheres
$v_i=\tu_i \cdot {\bf v}_i$.  This allows us to set up the equations of motion
that determine the time derivatives $\dot \phi_1$ and $\dot\phi_2$ as functions
of $\phi_1$ and $\phi_2$.  The problem of two coupled phase oscillators is
equivalent to the Kuramoto model \cite{Kuramoto84}, a classical model for
describing synchronization phenomena \cite{Pikovsky2001}.  Performing numerical
solutions to these dynamic equations, we find that for certain parameter
values, after long times the motion of both spheres becomes periodic with the
same frequency and the oscillations thus synchronize.  A synchronized state is
characterized by the phase lag $\delta=\left< \phi_2-\phi_1\right>$, where the
brackets denote a time average over one period. Regions of synchronous states
and a region of asynchronous states are indicated in the state diagram of
Fig.~\ref{fig_phasediag}b. The two regions of synchronous states correspond to
equal and opposite phases in the limit of infinite separation $r$. The values
of $\delta$ are displayed in Fig.~\ref{fig_phasevsbeta} as a function of the
angle $\beta$ for different distances $r=\vert {\bf x}_1^0-{\bf x}_2^0\vert$
between the two cilia. For some intervals of $\beta$, no synchronized solution
occurs and the spheres rotate in an asynchronous manner with two different
frequencies. Note that even if both cilia have identical properties, the
arrangement shown in Fig \ref{fig_phasediag}a breaks the symmetry and both
cilia can have different preferred frequencies when they interact.

The existence and stability of synchronous states can be studied analytically.
First, using Eq. (\ref{eq:friction0}), we find for large $r\gg a$
\begin{eqnarray}
  f_0&=&\tu_1 \cdot (\tensor \gamma (\ve{x}_1) +\kappa \tensor I) \ve{v}_1
   \nonumber -
   \tu_1 \cdot \tensor \gamma (\ve{x}_1) \tensor G(\ve{x}_2,\ve{x}_1)
   \tensor \gamma(\ve{x}_2) \ve{v}_2 \\&& +
  \mathcal{O}\left(\frac{f_0 a^2 D^4}{r^{6}}\right)
\end{eqnarray}
Here, we have used the fact that for large $r\gg a$, the interactions decay as
$G \propto r^{-3}$, while the friction terms scale with the sphere size,
$\gamma \propto a$.  Using this relation between the tangential velocities, we
can study synchronization. For an isolated sphere, the phase as a function of
time is denoted $\phi_0(t)$ (\ref{eq:fourier}). We denote by $\phi_1(t)$ 
the phase of a sphere 1 when interacting with sphere 2. The variation 
$\Delta T$ of the oscillation period $T$ of sphere 1
can then be written as
\begin{eqnarray}
  \label{eq:period-perturb}
  \Delta T&=&\int_{0}^{2 \pi} \left( \frac 1 {\dot \phi_1(\phi)} - \frac 1 {\dot
  \phi_0(\phi)} \right) d\phi 
=
\int_0^{T_0} \frac {v_0
  -v_1}{v_0} dt \nonumber \\
  &\approx &  -\frac 1 {f_0} \int_0^{T_0} \tu_1 \cdot \tensor \gamma (\ve{x}_1) \tensor
  G(\ve{x}_2,\ve{x}_1) \tensor\gamma (\ve{x}_2) \tu_2 \, v_2\, dt
\end{eqnarray}
where the last expression becomes correct in the limit of large $r/a$. In this
limit, the variation $\Delta T$ can be calculated using the Ansatz
$\ve{x}_1=\ve x_1^0 + \ve x (\phi_0(\omega t))$ and $\ve{x}_2=\ve x_2^0 + \ve x
(\phi_0(\omega t+\delta))$ resulting in a function $\Delta T(r,\beta,\delta)$.
A synchronized steady state exists if the change in period due to interactions
is the same for both spheres, i.e., $\Delta T (r,\beta,\delta)=\Delta
T(r,\beta+\pi,-\delta)$, where we have taken into account that exchanging both
spheres corresponds to $\beta\rightarrow \beta+\pi$ and $\delta \to -\delta$,
see Fig.~\ref{fig_phasediag}a.  Furthermore, a synchronized steady state is
locally stable if $\partial/\partial \delta[\Delta T (r,\beta,\delta)-\Delta
T(r,\beta+\pi,-\delta)]<0$.  In order to find explicit expressions for $\Delta
T$, we perform a systematic expansion in the small parameter $K$.  In the limit
of vanishing radius $a$, the friction matrix $\tensor \gamma$ becomes
independent of the height $z$. The velocity of a single sphere then becomes
constant along the trajectory ($K=0$).  As a consequence, the corresponding
variation $\Delta T_0$ of the period, resulting from
Eq.~(\ref{eq:period-perturb}), becomes symmetric with respect to $\delta\to
-\delta$ as well as with respect to $\beta\to \beta + \pi$ and the condition
that both particles have the same period is satisfied trivially, regardless of
the phase lag $\delta$. In order to find nontrivial synchronized states, we
have to go to first order in $K$
 where
 \begin{equation}
  \label{eq:secondcorrection}
  \Delta T = \Delta T_s - T_0 \frac {9\pi^2 \gamma_0 aABCDK}{{2(\gamma_0+\kappa)}\ell^2 r^3}
  \sin 2\beta \sin \delta 
\end{equation}
($\Delta T_s(\delta)\equiv \Delta T_s(-\delta)$ represents symmetric terms that
are irrelevant for synchronization).  At this order in $a$ and $r$, the
condition for a stable synchronized steady state is fulfilled with $\delta=0$
for $0<\beta<\pi/2$ and $\pi<\beta< 3\pi/2$ and with $\delta=\pi$ for
$\pi/2<\beta<\pi$ and $3\pi/2<\beta<2\pi$.  This is consistent with our
numerical solutions in the limit of large $r$, see Fig.~\ref{fig_phasevsbeta}.
For smaller distance $r$, the contribution of terms proportional $r^{-4}$
becomes relevant.

To conclude, we have shown that the hydrodynamic far field around a
periodically beating cilium has generic properties which do not depend on the
detailed beating pattern. Its main features can thus be generated by a sphere
moving on a tilted elliptical trajectory.  This captures the asymmetry of the
ciliary beat.  As a consequence, the resulting flow is a time periodic pattern
superimposed on a time-independent net flow.  If the beating cilium is
perturbed by external forces, this interferes with the internal force
generating process and affects the instantaneous angular velocity of
oscillations. We capture this effect by a linear relationship between external
force and the velocity $v$ of the sphere.  The force exerted on one cilium by
the hydrodynamic flow generated by the other couples both oscillators.  The
resulting synchronization phenomena depend on the distance $r$ between cilia
but also the parameters $a/D$ and $a/C$ which characterize the difference of
effective and recovery strokes.  No synchronization occurs if the motion is
rotationally symmetric or helical \cite{Kim.Powers2004,Reichert.Stark2005}.

A natural extension of our study is the generalization to a periodic lattice of
cilia attached to a surface.  Such situations correspond to some epithelial
cells which generate fluid flows on their surface and to microorganisms such as
paramecium.  In all these cases, the ciliary beat is organized in metachronal
waves which result from hydrodynamic and steric interactions between cilia.  In
different cells, these waves of ciliary strokes propagate along the surface in
different directions relative to the one defined by the working stroke
\cite{Machemer1972,Gheber.Priel1997}.  Other examples for hydrodynamically
stirred surfaces are realized in experiments where bacteria are attached to a
surface by their flagella which are driven by rotatory motors
\cite{Darnton.Berg2004}.  Each bacterium in this system could be represented by
a rotating sphere as in our description. In principle, the synchronization
effects between two rotating elements discussed here could be studied in such
experiments.  The extension of our study to many hydrodynamically coupled cilia
and the description of metachronal waves and other collective modes is left for
future work.

\begin{acknowledgments}
  This work was supported by the Slovenian Office of Science (Grants
  No.~J1-6502 and P1-0099).
\end{acknowledgments}


\begin{thebibliography}{16}
\expandafter\ifx\csname natexlab\endcsname\relax\def\natexlab#1{#1}\fi
\expandafter\ifx\csname bibnamefont\endcsname\relax
  \def\bibnamefont#1{#1}\fi
\expandafter\ifx\csname bibfnamefont\endcsname\relax
  \def\bibfnamefont#1{#1}\fi
\expandafter\ifx\csname citenamefont\endcsname\relax
  \def\citenamefont#1{#1}\fi
\expandafter\ifx\csname url\endcsname\relax
  \def\url#1{\texttt{#1}}\fi
\expandafter\ifx\csname urlprefix\endcsname\relax\def\urlprefix{URL }\fi
\providecommand{\bibinfo}[2]{#2}
\providecommand{\eprint}[2][]{\url{#2}}

\bibitem[{\citenamefont{Bray}(2001)}]{Bray2001}
\bibinfo{author}{\bibfnamefont{D.}~\bibnamefont{Bray}},
  \emph{\bibinfo{title}{Cell Movements}} (\bibinfo{publisher}{Garland Publ.,
  New York}, \bibinfo{year}{2001}), \bibinfo{edition}{2nd} ed.

\bibitem[{\citenamefont{Brennen and Winet}(1977)}]{Brennen.Winet1977}
\bibinfo{author}{\bibfnamefont{C.}~\bibnamefont{Brennen}} \bibnamefont{and}
  \bibinfo{author}{\bibfnamefont{H.}~\bibnamefont{Winet}},
  \bibinfo{journal}{Ann. Rev. Fluid Mech.} \textbf{\bibinfo{volume}{9}},
  \bibinfo{pages}{339} (\bibinfo{year}{1977}).

\bibitem[{\citenamefont{McGrath and Brueckner}(2003)}]{McGrath.Brueckner2003}
\bibinfo{author}{\bibfnamefont{J.}~\bibnamefont{McGrath}} \bibnamefont{and}
  \bibinfo{author}{\bibfnamefont{M.}~\bibnamefont{Brueckner}},
  \bibinfo{journal}{Curr.~Opin.~Genet.~Dev.} \textbf{\bibinfo{volume}{13}},
  \bibinfo{pages}{385} (\bibinfo{year}{2003}).

\bibitem[{\citenamefont{Cartwright et~al.}(2004)\citenamefont{Cartwright, Piro,
  and Tuval}}]{Cartwright.Tuval2004}
\bibinfo{author}{\bibfnamefont{J.~H.} \bibnamefont{Cartwright}},
  \bibinfo{author}{\bibfnamefont{O.}~\bibnamefont{Piro}}, \bibnamefont{and}
  \bibinfo{author}{\bibfnamefont{I.}~\bibnamefont{Tuval}},
  \bibinfo{journal}{Proc.\ Natl.\ Acad.\ Sci.\ USA}
  \textbf{\bibinfo{volume}{101}}, \bibinfo{pages}{7234} (\bibinfo{year}{2004}).

\bibitem[{\citenamefont{Okada et~al.}(2005)\citenamefont{Okada, Takeda, Tanaka,
  Belmonte, and Hirokawa}}]{Okada.Hirokawa2005}
\bibinfo{author}{\bibfnamefont{Y.}~\bibnamefont{Okada}},
  \bibinfo{author}{\bibfnamefont{S.}~\bibnamefont{Takeda}},
  \bibinfo{author}{\bibfnamefont{Y.}~\bibnamefont{Tanaka}},
  \bibinfo{author}{\bibfnamefont{J.~C.} \bibnamefont{Belmonte}},
  \bibnamefont{and} \bibinfo{author}{\bibfnamefont{N.}~\bibnamefont{Hirokawa}},
  \bibinfo{journal}{Cell} \textbf{\bibinfo{volume}{121}}, \bibinfo{pages}{633}
  (\bibinfo{year}{2005}).

\bibitem[{\citenamefont{Gueron and
  Levit-Gurevich}(1998)}]{Gueron.Levit-Gurevich1998}
\bibinfo{author}{\bibfnamefont{S.}~\bibnamefont{Gueron}} \bibnamefont{and}
  \bibinfo{author}{\bibfnamefont{K.}~\bibnamefont{Levit-Gurevich}},
  \bibinfo{journal}{Biophys.~J.} \textbf{\bibinfo{volume}{74}},
  \bibinfo{pages}{1658} (\bibinfo{year}{1998}).

\bibitem[{\citenamefont{Blake}(1971)}]{Blake.1971}
\bibinfo{author}{\bibfnamefont{J.~R.} \bibnamefont{Blake}},
  \bibinfo{journal}{Proc.~Camb.~Phil.~Soc.} \textbf{\bibinfo{volume}{70}},
  \bibinfo{pages}{303} (\bibinfo{year}{1971}).

\bibitem[{\citenamefont{Dufresne et~al.}(2000)\citenamefont{Dufresne, Squires,
  Brenner, and Grier}}]{Dufresne.Grier2000}
\bibinfo{author}{\bibfnamefont{E.~R.} \bibnamefont{Dufresne}},
  \bibinfo{author}{\bibfnamefont{T.~M.} \bibnamefont{Squires}},
  \bibinfo{author}{\bibfnamefont{M.~P.} \bibnamefont{Brenner}},
  \bibnamefont{and} \bibinfo{author}{\bibfnamefont{D.~G.} \bibnamefont{Grier}},
  \bibinfo{journal}{Phys.~Rev.~Lett.} \textbf{\bibinfo{volume}{85}},
  \bibinfo{pages}{3317} (\bibinfo{year}{2000}).

\bibitem[{\citenamefont{Lenz et~al.}(2003)\citenamefont{Lenz, Joanny, Julicher,
  and Prost}}]{Lenz.Prost2003}
\bibinfo{author}{\bibfnamefont{P.}~\bibnamefont{Lenz}},
  \bibinfo{author}{\bibfnamefont{J.~F.} \bibnamefont{Joanny}},
  \bibinfo{author}{\bibfnamefont{F.}~\bibnamefont{Julicher}}, \bibnamefont{and}
  \bibinfo{author}{\bibfnamefont{J.}~\bibnamefont{Prost}},
  \bibinfo{journal}{Phys.~Rev.~Lett.} \textbf{\bibinfo{volume}{91}},
  \bibinfo{pages}{108104} (\bibinfo{year}{2003}).

\bibitem[{\citenamefont{Kuramoto}(1984)}]{Kuramoto84}
\bibinfo{author}{\bibfnamefont{Y.}~\bibnamefont{Kuramoto}},
  \emph{\bibinfo{title}{Chemical Oscillations, Waves, and Turbulence}}
  (\bibinfo{publisher}{Springer}, \bibinfo{address}{Berlin},
  \bibinfo{year}{1984}).

\bibitem[{\citenamefont{Pikovsky et~al.}(2001)\citenamefont{Pikovsky,
  Rosenblum, and Kurths}}]{Pikovsky2001}
\bibinfo{author}{\bibfnamefont{A.}~\bibnamefont{Pikovsky}},
  \bibinfo{author}{\bibfnamefont{M.}~\bibnamefont{Rosenblum}},
  \bibnamefont{and} \bibinfo{author}{\bibfnamefont{J.}~\bibnamefont{Kurths}},
  \emph{\bibinfo{title}{Synchronization, {A} Universal Concept in Nonlinear
  Sciences}} (\bibinfo{publisher}{Cambridge Univeristy Press},
  \bibinfo{address}{Cambridge}, \bibinfo{year}{2001}).

\bibitem[{\citenamefont{Kim and Powers}(2004)}]{Kim.Powers2004}
\bibinfo{author}{\bibfnamefont{M.}~\bibnamefont{Kim}} \bibnamefont{and}
  \bibinfo{author}{\bibfnamefont{T.~R.} \bibnamefont{Powers}},
  \bibinfo{journal}{Phys.~Rev.~E} \textbf{\bibinfo{volume}{69}},
  \bibinfo{pages}{061910} (\bibinfo{year}{2004}).

\bibitem[{\citenamefont{Reichert and Stark}(2005)}]{Reichert.Stark2005}
\bibinfo{author}{\bibfnamefont{M.}~\bibnamefont{Reichert}} \bibnamefont{and}
  \bibinfo{author}{\bibfnamefont{H.}~\bibnamefont{Stark}},
  \bibinfo{journal}{Eur.\ Phys.\ J. E Soft Matter}
  \textbf{\bibinfo{volume}{17}}, \bibinfo{pages}{493} (\bibinfo{year}{2005}).

\bibitem[{\citenamefont{Machemer}(1972)}]{Machemer1972}
\bibinfo{author}{\bibfnamefont{H.}~\bibnamefont{Machemer}},
  \bibinfo{journal}{J. Exp.\ Biol.} \textbf{\bibinfo{volume}{57}},
  \bibinfo{pages}{239} (\bibinfo{year}{1972}).

\bibitem[{\citenamefont{Gheber and Priel}(1997)}]{Gheber.Priel1997}
\bibinfo{author}{\bibfnamefont{L.}~\bibnamefont{Gheber}} \bibnamefont{and}
  \bibinfo{author}{\bibfnamefont{Z.}~\bibnamefont{Priel}},
  \bibinfo{journal}{Biophys.~J.} \textbf{\bibinfo{volume}{72}},
  \bibinfo{pages}{449} (\bibinfo{year}{1997}).

\bibitem[{\citenamefont{Darnton et~al.}(2004)\citenamefont{Darnton, Turner,
  Breuer, and Berg}}]{Darnton.Berg2004}
\bibinfo{author}{\bibfnamefont{N.}~\bibnamefont{Darnton}},
  \bibinfo{author}{\bibfnamefont{L.}~\bibnamefont{Turner}},
  \bibinfo{author}{\bibfnamefont{K.}~\bibnamefont{Breuer}}, \bibnamefont{and}
  \bibinfo{author}{\bibfnamefont{H.~C.} \bibnamefont{Berg}},
  \bibinfo{journal}{Biophys.~J.} \textbf{\bibinfo{volume}{86}},
  \bibinfo{pages}{1863} (\bibinfo{year}{2004}).

\end{thebibliography}
\end{document}